\documentstyle[prl,aps]{revtex}
\begin{document}
\draft
\voffset=.5in
%@twocolumnfalse\endcsname

\title
{Reply to comment on Casimir energy for spherical boundaries}

\author {C. R. Hagen\cite{Hagen}}

\address
{Department of Physics and Astronomy\\
University of Rochester\\
Rochester, N.Y. 14627}

\maketitle

\begin{abstract}
It is shown that the recent criticism of Brevik {\em et al.} is in error.
\end{abstract}

In a recent work Brevik, Jensen, and Milton [1] have sought to rebut criticisms
raised by me of certain techniques used in the calculation of the Casimir
effect for a conducting sphere.  It is shown here that the points which they
raise are not in fact correct.

First, comment needs to be made upon the fact that ref [1] expresses great
concern that I have seen fit to question whether the Casimir effect has been
experimentally verified.  While recognizing my essential point that the
geometry of recent experiments [3] is different than that of the cases for
which the Casimir effect has been carefully calculated, they seek to assert
that the result is nonetheless verified at the level of a few per cent.  I
categorically reject the claim that any such verification has been achieved.
In order to establish the validity of recent experiments as constituting a
proof of the reality of the Casimir effect it is necessary to adopt the same
procedure as used in the parallel plate case--namely, the normal modes must be
calculated and a sum performed to get the vacuum energy in the presence of
appropriate boundaries.  This calculation has never been done for those cases
and thus it is far from clear what the significance is of data obtained in
recent experiments.

Going on to the issue of the mathematical points raised in ref. [1] I repeat
here the remark made by me in ref. [2] to the effect that outgoing wave
boundary conditions are {\em not} derivable in quantum field theory.  To
demonstrate the error in [1] where it is claimed that outgoing wave boundary
conditions can be derived from causal boundary conditions it needs only to be
observed that what is referred to in [1] as the third form of Eq.(1) is {\em
not} equal to the fourth form.  This is easily seen from the simple fact that
the discontinuity in the time derivative at $t=0$ of the third form is
proportional to a delta function in the spatial coordinates (in agreement with
the canonical commutation relations) whereas the fourth form is continuous at
$t=0$.  Thus the \lq\lq simple identity" which I am accused of failing to
appreciate
is manifestly incorrect.  In fact the very claim that causal and outgoing wave
boundary conditions are equivalent is one which must seem incomprehensible to a
practicing field theorist.  A final remark concerning this point has to do with
the fact that at the end of the Introduction of ref. [1] it is agreed that the
use of a large sphere \lq\lq in order to keep the eigenvalues real" is
\lq\lq not
incorrect".  This is in fact an admission that in ref. [1] and its antecedents
the eigenvalues are {\em not} real, a fact which seems quite odd in light
of the
fact that the Casimir effect is frequently described as originating in a shift
of the {\em real} energy eigenvalues of the normal modes of a system.

It is also alleged that my denial of a necessary connection between the
discontinuity of the stress tensor and the Casimir pressure is simply the
denial of the undeniable.  In Eq.(9) ref. [1] introduces an inhomogeneous term
in the expression for the four divergence of the stress tensor.  However, the
argument based on this introduction consists merely of rebuttal by a
definitional device.  Specifically, it is clear that so long as there is a
difference between the inside and outside stress there will be a term in the
divergence of the stress which is localized to the surface of the sphere.  This
is {\em defined} in [1] to be the force but certainly is not obviously equal to
the Casimir pressure, a question which can only be decided by explicit
calculation.  This has been done with considerable care in ref [2] with the
result that the Casimir pressure is {\em not} given by the discontinuity of the
stress tensor.  The response to this in [1] is that because I sometimes write
unregularized expressions in [2] it can be rejected out of hand.  I plead
guilty to the sin which is common to this subfield of occasionally delaying
explicit regularization until absolutely necessary.  However, if desired
one can
regularize from the outset with little difficulty.  What one then finds is that
$\alpha^{(i)}_{ln}$ goes to zero as $1/R$ times something independent of $R$.
Upon doing the (regularized) integrations and sums required one obtains the
result (as before) that the exterior modes make zero contribution in the limit
of large $R$.  It would seem that any remaining disagreements concerning this
claim should be resolved by citing specific alleged errors in my calculations
and a demonstration that the correction of such errors leads to results
consistent with the claims of ref. [1] concerning the stress tensor.  Indeed,
it has been found recently [4] that the failure of the relation between the
vacuum stress and the Casimir effect characterizes the parallel plate case as
well.  Since in this instance the required calculations are considerably
simpler than for the case of the sphere, it should be quite straightforward to
confirm or reject the claims of refs. [1] and [2].

Finally, there is the matter of improper contour manipulation which has been
questioned in [1].  While this matter is somewhat peripheral to the main issue,
some discussion of this aspect may be of interest.  The argument given in [1]
is essentially the same as that of ref. [2] of [1].  Since a detailed
explanation as to why that Wronskian based approach fails has already been
given in the appendix of [2], it would seem that the authors of ref. [1] want
to suggest the necessity of a direct refutation in the context of ref. [1].
While there are a number of problems associated with the derivation of [1] it
is simplest merely to point out that the error which has been shown to
characterize Eq.(1) is now merely repeated in the reversal of order of
integration in section II.

In conclusion I merely note that all the criticisms offered in ref. [2] remain
valid.  The suggestion in [1] to the effect that the stress tensor result
must be valid simply because it agrees with that obtained by other methods is
certainly a curious one, but nonetheless one which is manifestly contradicted
by direct calculation.

\acknowledgments

This work is supported in part by the U.S. Department of Energy Grant
No.DE-FG02-91ER40685.

\end{document}